\documentclass[prl,twocolumn,aps,amsmath,showkeys,showpacs,floatfix,nofootinbib]{revtex4-1}

\usepackage{graphicx}
\usepackage{dcolumn}
\usepackage{bm}
\usepackage{color}
\usepackage{bbold}

\newcommand{\ket}[1]{\left|#1\right\rangle} 
\newcommand{\bra}[1]{\left\langle#1\right|} 
\newcommand{\braket}[2]{\left< #1 \vphantom{#2} \right|
 \left. #2 \vphantom{#1} \right>} 
\newcommand{\op}[1]{\ensuremath{\Hat{\mathrm{#1}}}}

\def\vec#1{{\boldsymbol{#1}}}  

\begin{document}


\title{Non-Abelian spin singlet states of two-component Bose gases in artificial gauge fields}

\author{ T. Gra\ss$^1$, B. Juli\'a-D\'{\i}az$^{1}$,  N. Barber\'an$^2$, and M. Lewenstein$^{1,3}$}

\affiliation{$^1$ICFO-Institut de Ci\`encies Fot\`oniques, Parc Mediterrani 
de la Tecnologia, 08860 Barcelona, Spain}
\affiliation{$^2$Dept. ECM, Facultat de F\'isica, U. Barcelona, 08028 Barcelona, Spain}
\affiliation{$^3$ICREA-Instituci\'o Catalana de Recerca i Estudis Avan\c cats, 
08010 Barcelona, Spain}

\begin{abstract}
We study strongly correlated phases of a pseudo-spin-1/2 Bose gas in an
artificial gauge field using the exact diagonalization method. The atoms are
confined in two dimensions and interact via a two-body contact potential.
In Abelian gauge fields, pseudo-spin singlets are favored by pseudo-spin
independent interactions. We find a series of incompressible phases at fillings
$\nu=2k/3$. By comparison with the non-Abelian spin singlet (NASS) states,
constructed as zero-energy eigenstates of a $(k+1)$-body contact interaction, we
classify the non-trivial topology of the states. An additional spin-orbit
coupling is shown to switch between NASS-like states
and spin-polarized phases from the Read-Rezayi series.
\end{abstract}

\pacs{73.43.-f,03.65.Vf,37.10.Vz}
\keywords{Gauge fields. Ultracold Bose gases. Correlated states. Exact diagonalization.}
\maketitle

Topological phases of matter attract a great deal of attention since the late
1980s, when the fractional quantum Hall (FQH) effect of electrons in two
dimensions (2D) subjected to a strong perpendicular magnetic field was
recognized as a paradigm of such phases \cite{wen-niu}. It was then pointed out
that amongst these states there might be some which support non-Abelian
excitations, where the exchange of two quasiparticles is described by some
unitary matrix, as it shifts between different degenerate ground states
\cite{moore-read}. Recently, this property has triggered the interest in
topological phases, as quantum gates operating with non-Abelian anyons might
pave the way to fault-tolerant quantum computers \cite{nayak}. For this purpose,
the possibility of controlling or even designing quantum many-body states seems
to be advantageous if not inevitable. A very important tool to do so are cold
atoms which can, in various ways, be manipulated through interactions with light
\cite{mlbook}.
 Despite being electroneutral, the atoms may even couple to external gauge
potentials, which, instead of being true electromagnetic vector potentials, have
to be synthesized. Artificial magnetic fields have been achieved
experimentally by rotating the atomic cloud \cite{schweikhard} (see also
\cite{cooper-aip} for a review), though instabilities at high rotation
frequencies have so far prevented from reaching the highly
correlated regime. A promising alternative pursues the generation of
gauge fields by a laser dressing of the atoms
\cite{dalibard,*brunoPRA,*brunoNJP}. This method has already been proven
experimentally \cite{lin,*spielmanPRL}. It is readily generalized to non-Abelian
gauge fields \cite{fleischhauer}, and has recently allowed for synthesizing a
spin-orbit coupling \cite{spielman-sobec}.

 The strongly correlated states occurring in such systems have been numerically
investigated for the case of spinless or spin-polarized bosons
\cite{cooper-wilkin-gunn}, in which the incompressible phases are well described
by the Read-Rezayi (RR) series of parafermionic states \cite{read-rezayi},
including the Laughlin wave function \cite{laughlin}, and the Moore-Read (MR)
wave function \cite{moore-read}. Although also bosons with a pseudo-spin-1/2
degree of freedom have already been condensed
\cite{ketterle-2comp,*wieman-2comp}, and their behavior under rotation has been
studied both theoretically
\cite{mueller-ho-vortices,*reiman-1vortex,*ueda-vortices} and experimentally
\cite{cornell-vortices,*cornell-vortex-lattice}, few attention has yet been paid
to the strongly correlated regime in such systems. For the electronic FQH
effect, however, a variety of states have been proposed for systems with
spin, which despite the strong magnetic field might be in a singlet phase due to
the small gyromagnetic ratio of electrons in many solids. Within a
conformal field theory (CFT) framework, a generalization of the RR series to the
so-called non-Abelian spin singlet (NASS) series
\cite{ardonne-schoutens,*nass-nucl,*estienne-bernevig} has been formulated. Here
the term ``non-Abelian'' refers to the nature of the excitations. For
bosons, this series of topological states occurs at fillings $\nu=2k/3$, and
includes the Halperin state \cite{halperin} with Abelian excitations at $k=1$.
Like the RR states, the NASS states are constructed as the zero-energy
eigenstates of a repulsive $(k+1)$-body contact interaction.

In this paper we show, using exact diagonalization, that also two-body contact
(2BC) interaction leads to the formation of incompressible states precisely at
the NASS filling factors in a 2D Bose gas exposed to an Abelian gauge field. For
$k=1$, this is trivially the Halperin state, but also for $k=2$ significant
overlap with the corresponding NASS state and a similar spectral structure
indicate that the system can be brought into the NASS phase within this
realistic setup.

This finding is in close analogy to the spin-polarized case with RR-like states
\cite{cooper-wilkin-gunn} at filling $\nu=k/2$. However, as it is experimentally
difficult to get into the regime of low filling, the two-component system
benefits from supporting analogs of the RR states at higher filling.
Furthermore, its possibility of tuning independently inter-spin and intra-spin
interactions provides additional control.
For pseudo-spin-independent repulsive 2BC interactions, a mechanism is discussed
which favors pseudo-spin singlet states. We show that, counterintuitively,
states with zero interspin interaction energy are obtained by increasing the
interspin interaction strength.
Finally, effects stemming from a spin-orbit (SO) coupling can be integrated in
the framework of two-component gases by generalizing to a non-Abelian gauge
field. We investigate this scenario, and find a transition from the
single-component RR states
\cite{trombettoni,*trombettoniPRA,palmer-pachos,*komineas-cooper} to the NASS
states at SO coupling strengths corresponding to a Landau level (LL) degeneracy.

We consider a two-component Bose gas in 2D, which on the single-particle level
is described by the Hamiltonian $H_{\mathrm{sp}}=(\vec{p}-\vec{A})^2/(2m)$. The
vector potential $\vec{A}$ is given by
\begin{eqnarray}
\label{A}
 \vec{A} = B (0,x,0) \mathbb{1} + q (\sigma_x,\sigma_y,0).
\end{eqnarray}
The first term, acting equally on both pseudo-spin components, describes a
magnetic field of strength $B$ in the Landau gauge. The second term, with
$\sigma_i$ being the Pauli matrices, accounts for a Rashba SO coupling
controlled by the coupling strength $q$, which we set to zero in the first part
of this work. A detailed description how to implement this Hamiltonian for
electroneutral atoms can be found in the supplementary material of Ref.
\cite{trombettoni}.
As the boundary conditions we choose a periodic square of length $a$. This
eliminates edge effects from our study, and, moreover, generates a non-trivial
toroidal geometry, on which topological phases can be classified by the number
of degenerate ground states \cite{wen-niu}. Solving the single-particle
Hamiltonian yields, for $q=0$, the usual LL structure
\cite{yoshioka-halperin-lee}, where every state $\ket{n,j,s}$ within a fully
degenerate Landau level $n$ is characterized by a momentum quantum number $j$
and an additional two-fold degenerate spin quantum number
$s=\uparrow,\downarrow$. Since an energy gap $2B$ separates the different LLs,
we restrict ourselves to the lowest LL, $n=0$. Fixing the number of fluxes per
unit cell, $N_{\Phi}$, restricts $j$ to values running from 1 to $N_{\Phi}$, and
the many-body Hamiltonian is, up to a constant amounting for the kinetic energy, given by
\begin{eqnarray}
\label{H}
\op{H} = \sum_{\{j,s\}}  V_{\{j,s\}} \op{a}_{j_1s_1}^{\dagger} \op{a}_{j_2s_2}^{\dagger} \op{a}_{j_3s_3}\op{a}_{j_4s_4},
\end{eqnarray}
where $\op{a}_{js}$ annihilates a particle in state $\ket{0,j,s}$, and $\{j,s\}$ denotes the set of quantum numbers $j_1,\dots,j_4$ and $s_1,\dots,s_4$. The interaction conserves total momentum $J=\sum_i j_i \ \mathrm{mod} \ N_{\Phi}$, and the spin of the particles, so the coefficients read
\begin{eqnarray}
\label{V}
 V_{\{j,s\}} &=& \delta'_{j_1+j_2,j_3+j_4} \delta_{s_1,s_3} \delta_{s_2,s_4} \times \\ \nonumber && \bra{j_1,s_1}\otimes\bra{j_2,s_2} g_{s_1s_2} \delta(z_1-z_2) \ket{j_3,s_3}\otimes\ket{j_4,s_4},
\end{eqnarray}
with $g_{s_1s_2}$ the spin-dependent contact interaction strength of two
particles, and $\delta'$ a Kronecker delta modulo $N_{\Phi}$. The Hamiltonian
is diagonalized in blocks with a fixed particle number $N$, defining the filling
factor $\nu=N/N_{\Phi}$, fixed Haldane momentum $\vec{K}$ \cite{haldane}, and a
fixed spin polarization $S=N_{\uparrow}-N_{\downarrow}$, where $N_{\uparrow}$
($N_{\downarrow}$) denotes the number of spin-up (spin-down) particles.

\begin{figure}[t]
\includegraphics{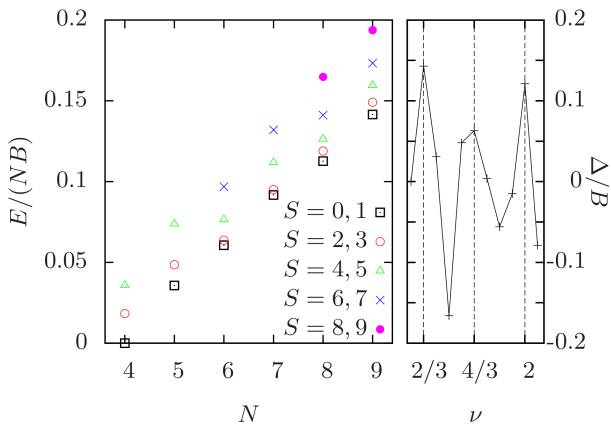}
\caption{\label{filling}
(Color online) Left: Ground state energies, not showing possible degeneracies,
at $N_{\Phi}=6$ for different $N$ and $S$ subspaces with $g_{s_1s_2}=g$. Even
(odd) values of $S$ correspond to even (odd) $N$. Right: $\Delta$ vs. $\nu$ for
$N_{\Phi}=6$ and $g_{s_1s_2}=g$.}
\end{figure}

First, we show that for spin-independent interactions, i.e. $g_{s_1s_2}=g$, the
singlet state is energetically favored.
This configuration is a natural choice, as many experiments 
\cite{ketterle-2comp,*wieman-2comp} work with the $\ket{F=1, m_F=0,1}$ states of
$^{23}$Na, or the $\ket{F=1,m_F=-1}$ and $\ket{F=2,m_F=1}$ states of $^{87}$Rb
with almost equal s-wave scattering lengths within and between the
components.
We start by contrasting the singlet to the fully polarized situation, $S=N$, in
which we recover the results from Ref. \cite{cooper-wilkin-gunn}, with
incompressible phases at $\nu=k/2$, forming the RR series. The fully polarized
zero-energy state with highest filling has $\nu=1/2$, and is the Laughlin state.
In the opposite limit of a fully unpolarized system, zero-energy states occur up
to $\nu=2/3$, where the 221-Halperin state is the exact unique ground state of
the 2BC interaction. The analytic expression of this state on a disc reads
\begin{eqnarray}
 \Psi_{\mathrm{H}} = \prod_{i<j}^{N_{\uparrow}} (z_{i \uparrow}-z_{j \uparrow})^2 \prod_{k<l}^{N_{\downarrow}} (z_{k \downarrow}-z_{l \downarrow})^2 \prod_{i,k} (z_{i \uparrow} - z_{k \downarrow}),
\end{eqnarray}
where we omit the irrelevant exponential term $\exp\left(-\sum
|z_i|^2/4\right)$. This wave function vanishes whenever two particles are at the
same position, and it is symmetric under exchange of two spin-up or two
spin-down particles. However, it is antisymmetric under exchange of particles
with different spin, and thus cannot be used for describing spinless bosons. For
this case, also the terms in the last product had to be squared, which would
result in a less dense Laughlin-like wave function. Generalizing this
observation, we state that any spatial wave function which solves the fully
polarized problem must also be a solution for $|S|<N$, while the opposite is not
true. Thus, for a spin-independent  interaction, it strictly follows for
the ground state energies $E(S_i)$ with $|S_1|<|S_2|$ that $E(S_1) \leq
E(S_2)$. 
 Our numerical data, partially shown in the left side of Fig. \ref{filling},
supports this finding. In fact, in all cases we have investigated, the ground
state of the $S=2$ subspace occurs in the spectrum of $S=0$ as a low-lying
excitation, except for $N=6$ and $N_{\Phi}=8$, where we have a degeneracy,
$E(S=0)=E(S=2)$. In most cases, thus, the ground state is a singlet, which is understood by noting that interactions of a
$\uparrow\downarrow$-pair can be fully avoided by a single zero of the form
$z_{i \uparrow} - z_{k \downarrow}$ in the wave function, while at least two
zeros must be spent if a pair of particles with equal spin is to be
anticorrelated in a similar way.

\begin{figure}[t]
\hspace{-0.8cm}
\includegraphics[width=9cm]{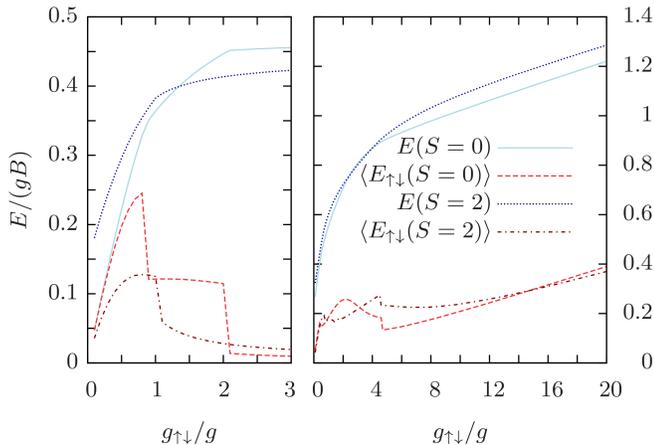}
\caption{\label{eud}
(Color online) Interaction energy for $N=6$ and $S=0,2$ as a function of $g_{\uparrow\downarrow}$. Left: $N_{\Phi}=6$. Right: $N_{\Phi}=5$.}
\end{figure}

Feshbach resonances allow to tune the spin-dependency of
interactions (cf. \cite{mlbook}). Increasing
$g_{\uparrow\downarrow}$ while leaving
$g_{\uparrow\uparrow}=g_{\downarrow\downarrow}=g$ should favor spin
polarizations with less $\uparrow \downarrow$-pairs. However,
this does not necessarily drive the system into the fully polarized
configuration: As shown in Fig. \ref{eud} for $N=6$ at $\nu=1$,
the energy of even the fully unpolarized system saturates for large
$g_{\uparrow\downarrow}$. In this limit, the real ground state has $S=2$ and
$E=0.449$ (in units of $gB$), close to the $S=0$ and $S=4$ ground states with
$E=0.466$ and $E=0.475$. All of them are well separated from the fully-polarized
ground state with $E=0.581$. In contrast to this, the energy of spin-unpolarized
systems will never saturate at larger filling, as is shown in Fig. \ref{eud} for
$\nu=6/5$. This different behavior is understood by noting that for $\nu=1$ the
number of available zeros in the wave function is sufficient to completely
suppress interactions between pairs of different spins, $\langle
E_{\uparrow\downarrow} \rangle \rightarrow 0$, while it is not for larger $\nu$.
Jumps in the curve of $\langle E_{\uparrow\downarrow} \rangle$ as a function of
$g_{\uparrow\downarrow}$ show that the states with $\langle
E_{\uparrow\downarrow} \rangle = 0$ are reached by several abrupt
re-organizations of the ground states.
We note that once the state with $\langle E_{\uparrow\downarrow}\rangle=0$ is
obtained, this state must be an eigenstate for arbitrary
$g_{\uparrow\downarrow}$. Thus, by abruptly switching off the interspin
interaction, one could produce binary mixtures of
highly entangled, non-interacting systems.

Coming back to spin-independent interactions, we ask now the question
whether, in analogy to the RR states for spin polarized systems, incompressible
states can be found amongst the spin-unpolarized ground states. Since
incompressibility is connected to a discontinuity in the chemical potential, we
define the particle-hole excitation gap as $\Delta(N) = N [ E(N+1)/(N+1) +
E(N-1)/(N-1) - 2 E(N)/N ]$, which is plotted for $N_{\Phi}=6$ in Fig.
\ref{filling} (right). Upwards peaks correspond to downward cusps
in the energy as a function of $\nu$, and signal incompressible
phases. They occur at fillings $\nu=2k/3$ with $k\in \mathbb{N}$. For these
fillings, a series of singlet states being exact zero-energy eigenstates of a
$(k+1)$-body contact interaction is known \cite{nass-nucl}. For $k=1$, this is
the 221-Halperin state, which has Abelian excitations, while the states for
$k>1$ are predicted to have non-Abelian excitations, which established the name
NASS series.

\begin{figure}[t]
\includegraphics{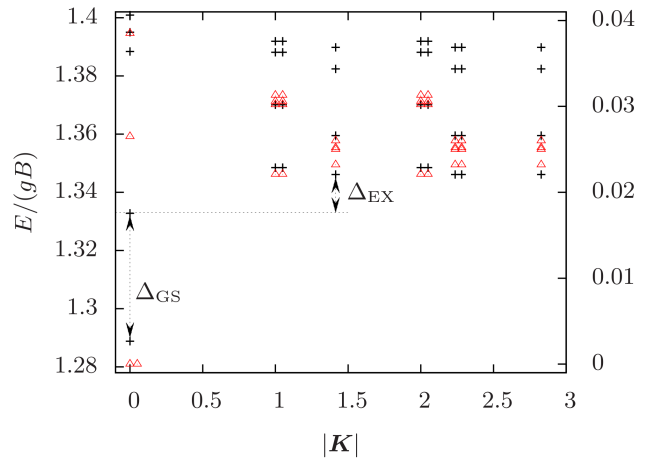}
\caption{(Color online) Energy spectra of singlet states for $N=12$ particles at filling $\nu=4/3$. Black crosses (with the left energy scale) show the results for an SU(2)-symmetric 2BC interaction, while red triangles (with the right energy scale) are the NASS spectrum for any three-body interaction with $g_{s_1s_2s_3}>0$. We do not show the three-fold center-of-mass degeneracy, but visualize further degeneracies by slight shifts in $\vec{K}$.
\label{2spectra}}
\end{figure}

We first construct these states for $k=2$ and 3 by diagonalizing the three- and
four-body contact interaction for up to $N=12$ particles: As predicted by CFT,
these states are characterized by a $(k+1)(k+2)/2$-fold topological degeneracy
on the torus. A subtlety occurs at $k=3$, where the CFT construction
\cite{nass-nucl}, being restricted to $N=2kp$ with $p \in \mathbb{N}$, is not
able to predict zero-energy ground states for all $N$ possible at $\nu=2$. In
fact, we find zero-energy states also for $N=8$ and 10, but, in contrast to the
states with $N=6$ and 12 which have the predicted ten-fold degeneracy, they are
non-degenerate, and thus belong to a different topological phase.

The overlaps of these states with the ground states of Eq. (\ref{H}), given in
Table \ref{ov}, turn out to be significant: For $k=2$, the states are exactly
equal for $N=4$, and even for a relatively large-sized system with $N=12$ the
overlaps are still around 0.8. Such a decrease of the overlap with large $N$
also occurs in the case of the RR states in single-component systems with 2BC
interaction \cite{cooper-aip}, suggesting the use of ground
state degeneracies to characterize the topological phase in the thermodynamic
limit \cite{wen-niu}. Topological degeneracies are usually lifted in finite
systems, as also happens here. In the case of $\nu=4/3$ we may consider the fact
that the six lowest eigenstates of $\op{H}$ agree in their Haldane momenta
$\vec{K}$ with the ones of the sixfold degenerate three-body eigenstates as a
signal for belonging to the same topological phase. Both spectra are shown for
$N=12$ in Fig. \ref{2spectra}. We note that both the degeneracy splitting
$\Delta_{\mathrm{GS}}$ and the gap to the excited band $\Delta_{\mathrm{EX}}$
behave non-monotonic with $N$. The situation is different for $\nu=2$, where the
candidates for the degenerate ground state manifold mix up with excited states. 

Our results therefore suggests that only the $\nu=4/3$-NASS phase can be
produced with 2BC interaction. For the experimental feasibility of this state to
be high, it should also be robust with respect to deviations from the
SU(2)-symmetric interaction. In the case of a $(k+1)$-body contact interaction
at $\nu=2k/3$, the numerical values of each interaction parameter do not
influence the ground states. Though this is different for 2BC interaction at
$\nu=4/3$, the overlaps with the NASS states remain high within a wide range of
$g_{\uparrow\downarrow}/g$, and have a maximum in the SU(2)-symmetric
configuration. For $N=8$, all overlaps are above 0.8 if
$g_{\uparrow\downarrow}/g \in [0.8,1.6].$

\begin{table}
\begin{center}
  \begin{tabular}{| l  c  c  c |c |l  c  c  c | }
   \cline{1-4}\cline{6-9}
    $N,N_{\Phi}$ & $\vec{K}$ & $|\braket{\mathrm{2b}}{\mathrm{3b}}|$ & $E_{2\mathrm{b}}$ &&
    $N,N_{\Phi}$ & $\vec{K}$ & $|\braket{\mathrm{2b}}{\mathrm{4b}}|$ & $E_{2\mathrm{b}}$ 
\\  \cline{1-4}\cline{6-9}  \cline{1-4}\cline{6-9}
     4,3  & (0,0)  & 1     & 0.435 && 12,6 & (0,0) & 0.334 & 2.604\\  
          &        & 1     & 0.520 &&       &       & 0.828 & 2.677\\ \cline{1-4}
     8,6  & (0,0)  & 0.906 & 0.901 &&       & (0,2) & 0.775 & 2.708\\
          &        & 0.920 & 0.908 &&       & (0,4) & 0.775 & 2.708\\ \cline{1-4}
     12,9 & (0,0)  & 0.729 & 1.289 &&       & (2,0) & 0.775 & 2.708\\
          &        & 0.885 & 1.333 &&       & (2,2) & 0.646 & 2.685\\
          &        &       &       &&       & (2,4) & 0.646 & 2.285\\
          &        &       &       &&       & (4,0) & 0.775 & 2.708\\
          &        &       &       &&       & (4,2) & 0.646 & 2.685\\
          &        &       &       &&       & (4,4) & 0.646 & 2.685\\
    \hline
  \end{tabular}
\end{center}
\caption{Overlaps between zero-energy eigenstates of $(k+1)$-body interaction with corresponding eigenstates of the 2BC interaction at energy $E_{2\mathrm{b}}$ (in units of $gB$) and Haldane momentum $\vec{K}$ (in units of $2\pi \hbar/a$). All states at filling $\nu=4/3$ have an additional three-fold center-of-mass degeneracy. \label{ov}}
\end{table}

Finally, we consider an additional SO coupling with $q>0$ in Eq. (\ref{A}).
This coupling mixes internal and external degrees of freedom on the
single-particle level, such that the
LLs become superpositions of the form $\ket{n,j,\pm} \equiv \alpha_{n}^{\pm}
\ket{n-1,j,\uparrow} + \beta_{n}^{\pm} \ket{n,j,\downarrow}$ where
$\alpha_{n}^{\pm}$ and $\beta_{n}^{\pm}$ are functions of $q$
\cite{trombettoni}. For general $q$, this lifts the previous spin-degeneracy,
as the  single-particle energies now read $E_{n}^{\pm} = 2Bn +2q^2 \pm
\sqrt{B^2+8Bq^2n}$. As has been pointed out in Refs.
\cite{trombettoni,palmer-pachos}, the system then behaves basically like a
single-component system, and the incompressible states are described by the RR
series projected into the lowest LL of the SO coupled system. However,
for $q^2/B = 2n+1$, a degeneracy between $E_{n}^-$ and $E_{n+1}^-$ occurs as the
single-particle ground state switches from $\ket{n,j,-}$ to $\ket{n+1,j,-}$. The
Hamiltonian from Eq. (\ref{H}) can then be adapted to the system with SO
coupling by identifying the $s$ quantum number with the two-fold degeneracy
between $n$ and $n+1$, and by re-defining $V_{\{js\}}$ accordingly. Note that in
this case $S$ does not correspond to the spin polarization, but quantifies the
population imbalance between $\ket{n,j,-}$  and $\ket{n+1,j,-}$. 

As observed in Ref. \cite{palmer-pachos} within a lowest LL approximation, the
projected RR states do not describe the system close to these degeneracy points.
For $q^2/B = 3$, the system becomes gapless at both the Laughlin and the MR
filling $\nu=1/2$ and 1. However, we find a clear gap in the spectrum at
$\nu=2/3$ with $N=6$, and a small gap at $\nu=4/3$ with $N=8$. 
Projecting this $\nu=2/3$ state into the lowest LL of the system without
SO coupling, and comparing with the 221-Halperin state, we find an overlap of
only 0.19. It should be noted that, in contrast to
the previous case of a pure spin degeneracy, interactions now can induce flips
within the two-fold degenerate manifold, and therefore $S$ is not conserved. For
a complete description of the state, we use generalizations of the Halperin
state, obtained by diagonalizing the 2BC interaction at $\nu=2/3$ in an Abelian
field ($q=0$) within different spin subspaces. Then the highest overlap
of 0.60 is obtained with the state having $S=-6$, which corresponds to all
particles being in $\ket{0,j,-}$. This is also the most populated LL, showing
that, despite the degeneracy on the single-particle level, interactions favor
the LL with lower $n$. The total fidelity for $-6
\leq S \leq 6$ is 0.82. For $\nu=4/3$ with $N=8$,
but now comparing the non-degenerate ground state of the SO coupled
system with the ground states of a three-body contact interaction at
$q=0$ for different $S$, we still obtain a total fidelity of 0.52, summing from
$-8 \leq S \leq 8$. 

We have investigated two-component Bose gases in  artificial gauge fields. The
possibility of tuning the interaction constants allows to control and manipulate
the strongly correlated nature of the eigenstates. Our main finding is
the formation of incompressible phases described by the NASS series within the
experimentally feasible setup of particles interacting via two-body contact
interaction. The internal degree of freedom permits also to
realize systems with spin-orbit coupling.

This work has been supported by 
EU (NAMEQUAM, AQUTE), ERC (QUAGATUA), 
Spanish MINCIN (FIS2008-00784, FIS2010-16185, 
FIS2008-01661, 2009 SGR1289), Alexander von Humboldt Stiftung, and AAII-Hubbard. 
B.~J.-D. is supported by the Ram\'on y Cajal program.
M. L. acknowledges Hamburg Theory Award.

\textit{Note added:} After submission of this article, a preprint by S. Furukawa
and M. Ueda appeared presenting additional numerical evidences for the NASS
phase \cite{furukawa-ueda}.

\bibliography{lett.bib}
\end{document}